\documentclass[12pt]{article}
\usepackage[margin=1in]{geometry}
\date{}
\usepackage{amssymb,amsmath}
\usepackage{bm}
\usepackage{natbib}
\usepackage{color}
\usepackage{multirow}

\newtheorem{thm}{Theorem}
\newtheorem{unnumber}{}

\newtheorem{proof}[unnumber]{Proof}
\newenvironment{pf}{\begin{proof}\normalfont}{\end{proof}}
\newtheorem{represent}[unnumber]{Representation}
\newenvironment{repres}{\begin{represent}\normalfont}{\end{represent}}
\newcommand{\qed}{\quad$\Box$}
\newtheorem{example}{Example}[section]
\newenvironment{eg}{\begin{example}\normalfont}{\end{example}}
\newtheorem{egcontinued}[unnumber]{Example~\ref{eg:kempton} continued}

\title{The Square Array Design}

\begin{document}

\author{R.~A.~Bailey\thanks{School of Mathematics and Statistics,
  University of St Andrews,
  St Andrews KY16 9SS, UK. Email: rab24@st-andrews.ac.uk
  Orchid: 0000-0002-8990-2099}
  \and L.~M.~Haines\thanks{Department of Statistical Sciences, University of Cape Town, 
    Private Bag 7700, Rondebosch South Africa.
    Corresponding author: E-mail: linda.haines@uct.ac.za~
Orchid: 0000-0002-8843-5353 }}

\maketitle

\begin{abstract}
 This paper is about the construction of augmented row-column designs
  for unreplicated trials.  The method uses the representation of a
  $k \times t$ equireplicate incomplete-block design with $t$~treatments
  in $t$~blocks of size~$k$, termed an auxiliary block design, as a
  $t \times t$ square array design with $k$ controls, where $k<t$. 
  This  can  be regarded as an extension of the representation of a 
  Youden square as a partial latin square for unreplicated
  trials.
  Properties of the designs, in particular in relation
  to connectedness and randomization, are explored. Particular attention is
  given to square array designs which minimize the average  variances of the
  estimates of paired comparisons between test lines  and controls and between
  test-line and test-line effects. The use of equireplicate cyclic designs as
  auxiliary block designs is highlighted. These provide a flexible and workable
  family of augmented row-column square array designs.
Designs whose auxiliary block designs are not cyclic are also covered.
\end{abstract}


\section{Introduction}
\label{sec:intro}
The construction of designs for plant breeding experiments is challenging
because the quantity of seed available for each test line is usually sufficient
only  for a single plot. 
Augmented designs which
incorporate controls and thereby facilitate comparisons between the
unreplicated test lines are well-researched and widely used. The experiments
are most commonly implemented as block designs, with either complete or
incomplete blocks, so that the blocks accommodate one-way heterogeneity in the
field \citep{piephoag:16}. Augmented row-column designs which  allow for
two-way heterogeneity are not as straightforward to construct
and have received limited attention in the literature.
The early papers of \citet{fedr:75}, \citet{fnr:75}, \citet{linp:83}
and  \citet{williamsj:03} provide somewhat restrictive augmented row-column
designs, while the more recent papers by \citet{piephoag:16} and
\citet{vontp:20} are primarily concerned with the introduction of an
additional blocking structure in order to accommodate regional heterogeneity. 

The present study is concerned with a family of augmented row-column designs
for unreplicated trials which are theoretically tractable and 
practically appealing. The paper is organised as follows. The designs of
interest, that is the  $t \times t$ square array designs with $k$ controls,
where $2<k<t$, and the attendant fixed-effects model adopted for analysis, are
introduced in Section~\ref{sec:prelim}. The main results relating to the
construction and randomization of augmented row-column square array designs
developed from
auxiliary block designs are presented formally in Section~\ref{sec:main}.
Section~\ref{sec:cyc} provides a comprehensive account of  the use of
equireplicate cyclic designs as auxiliary block designs.
Section~\ref{sec:other} 
extends this to auxiliary block designs which are not cyclic
and provide augmented square array designs with low values 
of the average variance metrics. 
A brief discussion of the
paper and some  pointers for future research are given in Section~\ref{sec:end}.
An R package for constructing auxiliary block designs  and square array 
designs and  R programs to reproduce the examples in the paper are available 
in the Supplementary Material.  

\section{Preliminaries}
\label{sec:prelim}

The designs for  unreplicated trials introduced here are augmented row-column
designs   based on square arrays. Specifically, consider a row-column design
with plots arranged in $t$~rows and $t$~columns and  with $k$ controls
  (with $k<t$)
located once in each of the rows and once in each of the columns. There are
thus a total of  $t k$ plots for the controls and $t (t-k)$ plots for the test
lines. The designs are implemented as follows. 
Initially, the $k$~controls are randomly allocated to the specified sets of
plots occurring once in each row and once in each column, and the test lines
are randomly allocated to the remaining plots.  
Then, following  \citet[p.~108]{rab:08}, the rows and columns of the square
array design are randomly permuted.

A fixed effects model for the $t^2$~yields of the design and a total of
$v=k+t(t-k)$  treatments, that is controls and test lines, is adopted.
Specifically, the  $t^2 \times 1$ vector of yields, denoted $\mathbf{y}$, is expressed
as 
\[ 
\mathbf{y} = \mathbf{X} \bm{\tau} + \mathbf{Z}_r \bm{\rho} +
\mathbf{Z}_c \bm{\gamma} +\mathbf{e},
\]
where  $\bm{\tau}$ is a $v \times 1$ vector of fixed treatment effects with a
$t^2 \times v$  design matrix~$\mathbf{X}$, and $\bm{\rho}$ and $\bm{\gamma}$
are vectors
representing fixed effects for the $t$~rows and $t$~columns with attendant
$t^2 \times t$ design matrices~$\mathbf{Z}_r$ and~$\mathbf{Z}_c$, respectively.
The error term~$\mathbf{e}$
is assumed to be distributed as $N(0,\sigma^2 \mathbf{I})$,
where $\mathbf{I}$~is the identity matrix,
and the variance $\sigma^2$ is taken, without loss of generality, to be one.
The treatment information matrix associated with the model follows immediately
and is given by
$\mathbf{C} = \mathbf{X}^\top(\mathbf{I}
- \mathbf{Z}(\mathbf{Z}^\top \mathbf{Z})^{-} \mathbf{Z}^\top)\mathbf{X}$, where
  $\mathbf{Z}=[\mathbf{Z}_r ~ \mathbf{Z}_c]$
and $(\mathbf{Z}^\top \mathbf{Z})^{-}$ is a generalized inverse of
$\mathbf{Z}^\top \mathbf{Z}$. Then, if $\mathbf{C}$~has
rank $v-1$, the design is connected and all treatment contrasts are estimable.

The precision with which contrasts of the treatment effects are estimated
is assessed by 
the average  variance of estimates of the pairwise differences between the
test-line effects, denoted $A_{tt}$, between the control and test-line effects,
denoted $A_{ct}$, and between the control effects, denoted $A_{cc}$.  
These metrics can
be invoked as criteria for comparing candidate
designs, with $A_{tt}$ deemed to be of primary interest and $A_{ct}$ of
secondary interest. The metric $A_{cc}$ is introduced here for completeness.
Thus, $t \times t$ square array designs for which $A_{tt}$~is a minimum and
$A_{ct}$~is small, and possibly a minimum, are sought.
Section~\ref{sec:metrics} shows 
that these two metrics achieve their minima together.

\citet{kem:84} suggested a range of $20$ to $25$\% for the percentage  of 
plots  occupied by the  controls  in unreplicated trials
as a balance between theory and practice. This rule-of-thumb remains in 
general use today and is broadly followed in the examples
in the present  paper.
In addition, the number of error degrees of freedom  in a square array design is equal to
  \[tk-2(t-1) -(k-1) -1 = (t-1)(k-2),\] which is zero if $k=2$.  The assumption 
that $k\geq 3$ is therefore made  throughout the text.

An early example of an augmented row-column square array design
was given by \citet[pp.~33--34]{fedr:75}. They  transformed a $3 \times 7$ Youden
square  by interchanging rows and letters into a $7 \times 7$ augmented
row-column design with three controls; the construction is clearly
illustrated in their paper.
It seems that the Youden square has been interpreted as the key component
in this construction and that, as a consequence, the approach has not been
pursued. Indeed, \citet{vontp:20} state that ``$\ldots$ the dimensions are
restrictive since they [sic] used Youden designs.'' This example forms the
basis for a much broader strategy of construction for augmented row-column
square array designs from equireplicate incomplete-block designs which is
developed in the present study.  

\section{Main results}
\label{sec:main}

\subsection{Auxiliary block designs}
\label{sec:aux}
The interpretation of the Youden square as a balanced incomplete-block design
(BIBD) and,  more broadly, as an equireplicate, incomplete-block design in the
construction of \citet{fedr:75} can be used to great advantage by
noting the following theorem in \citet[p.~235]{rab:08}.

\begin{thm}
  Suppose that $\Gamma(k,t)$ is an equireplicate incomplete-block design for
  $t$~treatments in $t$~blocks of size~$k$. Then the design can be presented
  as a $k \times t$ rectangle, where the columns are labelled by the blocks,
  the entries in each column are the $k$~treatments in the relevant block,
  and each treatment occurs once in each row. 
\label{thm:marriage}
\end{thm}

It is then straightforward to show, following  \citet{fedr:75}, that
interchanging the rows and letters of a $k \times t$ rectangular design
generated from an equireplicate incomplete-block design $\Gamma(k,t)$  yields
an augmented row-column $t \times t$ square array design with $k$ controls.
The design $\Gamma(k,t)$
is therefore termed the
\textit{auxiliary block design} for the resultant augmented row-column square
array design.
\citet{mukerjee:24} introduced
the concept of a \textit{primal} as the block design for the controls alone in
a design for unreplicated trials which accommodates one-way heterogeneity. The
notion of an auxiliary block design introduced here can be construed as being
similar to that of a primal but differs in that the design also specifies the
coordinates of the controls in the   design with test lines.    

For ease of interpretation here, the representation of the
$k \times t$  rectangle as a $t \times t$ square array design with $k$~controls,
denoted $\Sigma(t,k)$, uses  a minor modification of that in
\citet{fedr:75}.
The numbered
treatments in the $j$th column of the $k \times t$ auxiliary block \mbox{design} 
specify the column  coordinates of the controls in the $j$th~row of the
\mbox{resultant} $t \times t$ square array design with~$k$ controls,
for $j=1, \ldots, t$. To illustrate, the transformation of the
$3 \times 12$  auxiliary block design introduced as an equi\-replicate
incomplete-block design  by \citet{baileys:86} into a
$12 \times 12$ augmented row-column square array design with $3$~controls
is shown in Figure 1. 
The representation can be stated formally as follows and is used
throughout the text. 
\begin{figure}[hbt!]
\caption{ Representation of (a)~a $3\times 12$ auxiliary block design taken from Bailey and Speed (1986)
   and (b)~the resultant $12 \times 12$ square array
    design with $3$~controls.\label{fig:sbdes}}
\begin{center}
  $ \begin{array}{r|c|c|c|c|c|c|c|c|c|c|c|c|}
      \multicolumn{1}{c}{}  & \multicolumn{1}{c}{1} & \multicolumn{1}{c}{2} &
      \multicolumn{1}{c}{3} & \multicolumn{1}{c}{4} & \multicolumn{1}{c}{5}
      & \multicolumn{1}{c}{6} & \multicolumn{1}{c}{7} & \multicolumn{1}{c}{8}
      & \multicolumn{1}{c}{9} & \multicolumn{1}{c}{10} &
      \multicolumn{1}{c}{11} & \multicolumn{1}{c}{12}\\
      \cline{2-13}
A & 2 & 4 & 9 & 12 & 11 & 5 & 6 & 7 & 1 & 10 &3 & 8\\
      \cline{2-13}
\mbox{(a)}\qquad      B & 3 & 5 & 8 & 11 & 1 & 12 & 9 & 10 & 7 & 2 & 4 & 6\\
      \cline{2-13}
      C & 1 & 6 & 7 & 10 & 8 & 2 & 3 & 4 & 5 & 9 &  11 & 12\\
      \cline{2-13}
\multicolumn{1}{c}{}\\
\multicolumn{1}{c}{}   & \multicolumn{1}{c}{1} & \multicolumn{1}{c}{2} &
      \multicolumn{1}{c}{3} & \multicolumn{1}{c}{4} & \multicolumn{1}{c}{5}
      & \multicolumn{1}{c}{6} & \multicolumn{1}{c}{7} & \multicolumn{1}{c}{8}
      & \multicolumn{1}{c}{9} & \multicolumn{1}{c}{10} &
      \multicolumn{1}{c}{11} & \multicolumn{1}{c}{12}\\
      \cline{2-13}
      1  & C & A & B & & & & & & & & &\\
      \cline{2-13}
      2 & & & & A & B & C & & & & & &\\
      \cline{2-13}
      3 & & & & & & & C & B & A & & &\\
      \cline{2-13}
      4 & & & & & & & & & & C & B & A\\
      \cline{2-13}
      5 & B & & & & & & & C & && A  &\\
      \cline{2-13}
      \mbox{(b)} \qquad 6 & & C & & & A & & & & & & & B\\
      \cline{2-13}
      7 & & & C & & & A & & & B & & &\\
      \cline{2-13}
      8 & & & & C & & & A & & & B & &\\
      \cline{2-13}
      9 & A & & & & C & & B & & & & &\\
      \cline{2-13}
      10 & &  B & & & & & & &  C & A & &\\
\cline{2-13}
11 & & & A & B & & & & &  & & C &\\
\cline{2-13}
12 & & & & & & B & & A & & & & C\\
\cline{2-13}
      \end{array}$    
\end{center}

\end{figure}
\begin{repres}
  If the  $(i,j)$-th entry in the $k \times t$ rectangular design  has integer
  symbol $s$, then the $(j,s)$-th entry in the square array design is given by
  $i$, where $i=1$, \ldots, $k$, $j=1$, \ldots, $t$ and $s =1$, \ldots, $t$.
\end{repres}
The representation of the Youden square as a
square array design precedes that of \citet{fedr:75}.
It was reported  by \citet{fisher:38} and coincides with that
given here.

The following theorem shows that the metric $A_{cc}$ does not depend on the choice
of auxiliary block design.

\begin{thm}
  Consider a $t \times t$ square array design with $k$~controls, where $k<t$.
  Then, irrespective of the choice of auxiliary block design $\Gamma(k,t)$,
  the average  metric for comparing of controls is given by $A_{cc} = 2/t$.
\label{thm:acc}
\end{thm}

\begin{pf}
  Suppose that $i$ and $j$ are control treatments. Since each control treatment
  occurs exactly once in each row and once in each column, the best linear
  unbiased estimator (BLUE) of the difference $\tau_i -\tau_j$ is given by the
  difference between the mean yields on those treatments. Since all control
  treatments have replication $t$, this has variance~$2/t$. \qed
  \end{pf}

\subsection{Relationship between metrics}
\label{sec:metrics}

\begin{thm}
\label{thm:attact}
  Consider a square array design as in Theorem~\ref{thm:acc}, with $k$ controls
  and $t_1$ test lines, where $t_1=t(t-k)$ and $v=k+t_1$.  Then
  \[
  A_{ct} = \frac{k-1}{kt} +\frac{1}{k(t-k)} +\frac{(t_1-1)}{2t_1}A_{tt}.
  \]
\end{thm}

\begin{pf}
  Consider the treatment  information matrix $\mathbf{C}$ and its Moore-Penrose
  inverse $\mathbf{C}^-$.  Number the controls $1$, \ldots, $k$ and the test lines
  $k+1$, \ldots, $k+t_1$.

  Since each control treatment occurs exactly once in each row and in each
  column, the values of $\rho_i$ and $\gamma_j$ can be constrained so that
  $\rho_1 + \cdots + \rho_t = \gamma_1 + \cdots + \gamma_t=0$.  If $\ell$ is a
  test line then the yield on the single plot containing this gives the only
  information about $\tau_\ell$.   The BLUEs of the constrained row and
  column coefficients can be obtained from the control data.  Since the
  contrasts between control treatments are orthogonal to rows and to columns,
  the variance of the BLUE of $\tau_\ell -\tau_m$ is the same for all
  control treatments~$m$.

  The proof of Theorem~\ref{thm:acc} shows that, if $i$ and $j$ are control treatments,
  the variance of the BLUE of $\tau_i-\tau_j$ is $2/t$.  Let $m$ be
  a third treatment, which may be either a control treatment or a test line.
  Then the above statements show that
  $
  C^-_{ii} + C^-_{mm} -2C^-_{im} =   C^-_{jj} + C^-_{mm} -2C^-_{jm} 
  $,
  and so
  \begin{equation}
    C^-_{ii} - C^-_{jj} = 2(C^-_{im}- C^-_{jm}).
    \label{eq:CC}
  \end{equation}
  Since $C^-_{ij} = C^-_{ji}$ and all row-sums of $\mathbf{C}^-$ are zero,
  Equation~(\ref{eq:CC}) shows that $C^-_{ii} = C^-_{jj}$ when $i$ and $j$
  are both controls.  Denote this constant by $f$.  Then $C^-_{ij} = f - 1/t$,
  which is denoted here by $g$.

  The sum of the entries in every row of $\mathbf{C}^-$ is zero, so the sum
  of the last $t_1$ entries in any control row is $h$, where
 \begin{equation}
    f + g(k-1) + h =0.
    \label{eq:fgh}
    \end{equation}
  Using the fact that $f-g=1/t$ gives $ft + (ft-1)(k-1) + ht=0$, and so
  \begin{equation}
fkt=k-1-ht.
    \label{eq:fh}
  \end{equation}

    The sum of the entries in the top-right $k \times t_1$ corner of $\mathbf{C}^-$
  is $kh$.
  All column sums of $\mathbf{C}^-$ are zero, so the sum of the entries in the
  bottom $t_1\times t_1$ corner is equal to $-kh$.
  The contrast between all the control treatments and all the test lines is orthogonal
  to rows and to columns, so its variance is $1/kt+1/t_1$.  The
  vector~$\mathbf{v}$ for
  this contrast has $k$ entries equal to $1/k$ then the remaining
  $t_1$ entries equal to $-1/t_1$.  Thus
  \[
  \frac{1}{tk} + \frac{1}{t_1} = \mathbf{v}^T \mathbf{C}^- \mathbf{v} =
  \frac{kf+gk(k-1)}{k^2} - \frac{kh}{t_1^2} -\frac{2hk}{kt_1}
  = -\frac{kh}{k^2}- \frac{kh}{t_1^2} -\frac{2hk}{kt_1}
  \]
  from Equation~(\ref{eq:fgh}).  Multiplying both sides by $kt_1^2$ gives
  $
  t_1t =-h(k+t_1)^2 
  $.

  Let $d=\sum_{i=k+1}^v C^-_{ii}$.  Then
    \begin{equation}
  \frac{t_1(t_1-1)}{2}A_{tt} = (t_1-1)d -(-kh-d) = t_1d+ kh.
\label{eq:long}
    \end{equation}
  (The left-hand side of this agrees with Equation (4.2) in \citet{fnr:75}.)

  Finally, all four corners of $\mathbf{C}^-$ can be used to
  calculate $A_{ct}$.  This gives $kt_1 A_{ct} = t_1kf + kd -2kh$.  Hence
 \begin{eqnarray*}
    A_{ct} & = & f + \frac{d}{t_1} -\frac{2h}{t_1}\\
    & = & \frac{(k-1)}{kt}-\frac{h}{k}+ \frac{(t_1-1)}{2t_1}A_{tt}
    -\frac{kh}{t_1^2}
    -\frac{2h}{t_1},
    \quad \mbox{from Equations~(\ref{eq:fh}) and (\ref{eq:long}),}\\
    & = & \frac{(k-1)}{kt}     +\frac{(t_1-1)}{2t_1}A_{tt}-
    \frac{h}{t_1^2k}(t_1^2 +k^2 +2t_1k)\\
      & = &   \frac{(k-1)}{kt} + \frac{(t_1-1)}{2t_1}A_{tt} +
    \frac{t}{kt_1},
\\      & = &   \frac{(k-1)}{kt} + \frac{(t_1-1)}{2t_1}A_{tt} +
    \frac{1}{k(t-k)}. \quad \mbox{\qed}
 \end{eqnarray*}
\end{pf}

Theorem~\ref{thm:attact} shows that the metrics $A_{tt}$ amd $A_{ct}$ are positive
linear functions of each other.
Denote by $A_{abd}$ the average variance of the estimates of pairwise
  differences of the treatment effects in the  auxiliary block design when it is used
  in its usual setting.  
Williams and Piepho (2025) have invoked results from the paper by Patterson and Williams (1976) to prove that
 \begin{equation}
 A_{tt} =  2+ \frac{2 t (t-1)}{t_1-1} \bigg(A_{abd}-\frac{2}{t}\bigg).
 \label{eq:att}
\end{equation}
It thus  follows from Theorem~\ref{thm:attact}  that  both $A_{tt}$ and $A_{ct}$ are  positive linear functions of $A_{abd}$ and that
\begin{equation}
 A_{ct} =  1 + \frac{1}{t} + \frac{t-1}{t-k} \bigg(A_{abd}-\frac{2}{t}\bigg).
\label{eq:act}  
\end{equation}
These results were in fact developed  as a proof of concept early in the present study, with patterns in the average variance metrics of cyclic square array designs identified numerically, but were not reported at the time. The expressions for $A_{ct}$ and $A_{tt}$  resonate  with the general form of the  metrics in unreplicated trials.

\subsection{Randomization}
\label{sec:rand}

The $t \times t$ square array designs with $k$~controls introduced here
are row-column designs,  and randomization can  be implemented by randomly
permuting the rows and, independently,  the columns of the array
\citep[p.~108]{rab:08}. Such a randomization procedure should, however, be
strongly valid in order to avoid bias in the estimates of treatment contrasts
involving the test lines. This can be achieved by taking permutations from any
of the doubly transitive subgroups of the symmetric group $S_t$, with the
choice of subgroup not important \citep{gh:50, bailey:83}.  

For a $t \times t$ square array design with $t$~prime, the
doubly transitive subgroup of~$S_t$ comprising permutations of the form
$x \mapsto a x + b$, where $a$ and $b$ are integers modulo~$t$  and $a \ne 0$,
is computationally tractable  and the requisite permutations can be drawn
randomly. If $t$~is a power of a prime, the analogous permutations give a
doubly transitive subgroup of~$S_t$ if the items being permuted are labelled by
the elements of the finite field of order~$t$. For all other values of~$t$,
doubly transitive subgroups of the symmetric group~$S_t$ can be identified
either from specific results in group theory or from a programming language
such as \citet{gap}. 

A powerful theorem
about this randomization is now introduced. 

\begin{thm}
\label{th:rand}
  The values of average variance metrics $A_{tt}$ and $A_{ct}$ for a
  $t \times t$ square array design with $k$~controls depend only on the
  attendant auxiliary block design $\Gamma(k,t)$. They are therefore not
  altered by randomization of rows or randomization of columns.
  \end{thm}

\begin{pf}
  Project the vector of yields~$\mathbf{y}$ onto the orthogonal complement of
  the subspace spanned by row- and column-vectors. This is achieved by
  pre-multiplying~$\mathbf{y}$ by the matrix
  $(\mathbf{I} -t^{-1} \mathbf{Z}_r \mathbf{Z}_r^\top)(\mathbf{I}
  -t^{-1}\mathbf{Z}_c \mathbf{Z}_c^\top)$.
  Since $t^{-2} \mathbf{Z}_r (\mathbf{Z}_r^\top \mathbf{Z}_c)\mathbf{Z}_c^\top =
  t^{-2} \mathbf{Z}_r \mathbf{J}_t \mathbf{Z}_c^\top =
  t^{-2} \mathbf{J}_{t^2}$, this means pre-multiplication by 
  $\mathbf{I} -t^{-1} \mathbf{Z}_r \mathbf{Z}_r^\top -t^{-1}\mathbf{Z}_c \mathbf{Z}_c^\top
  + t^{-2} \mathbf{J}_{t^2}$. 
The information matrix for all treatments, $\mathbf{C}$, is therefore given by
\begin{eqnarray*}
  \mathbf{C} &=& \mathbf{X}^\top (\mathbf{I}
  -t^{-1} \mathbf{Z}_r \mathbf{Z}_r^\top-t^{-1} \mathbf{Z}_c \mathbf{Z}_c^\top +
  t^{-2} \mathbf{J}_{t^2})\mathbf{X} \\
  &=& \mathbf{X}^\top \mathbf{X}
  - t^{-1} \mathbf{X}^\top \mathbf{Z}_r \mathbf{Z}_r^\top \mathbf{X}
  -t^{-1} \mathbf{X}^\top \mathbf{Z}_c \mathbf{Z}_c^\top \mathbf{X}
+ t^{-2} \mathbf{X}^\top \mathbf{J}_{t^2} \mathbf{X}\\
&=& \mathbf{X}^\top \mathbf{X}-t^{-1} \bm{\Lambda}_r -t^{-1}\bm{\Lambda}_c
+t^{-2}\mathbf{X}^\top \mathbf{J}_{t^2} \mathbf{X}.
\end{eqnarray*}
Here $\mathbf{X}^\top \mathbf{X}$ is the diagonal matrix of treatment replications,
$\bm{\Lambda}_r$~is the concurrence matrix for treatments in rows,
$\bm{\Lambda}_c$~is the concurrence matrix for treatments in columns, and the
$(i,j)$-th entry of $\mathbf{X}^\top \mathbf{J}_{t^2} \mathbf{X}$ is equal to the product
of the replications of treatments $i$ and~$j$. None of these quantities is changed
by permutations of rows and columns. \qed
  \end{pf}

Thus, 
the linear combinations of responses which
give the best linear unbiased estimators of any difference between test lines,
or between a test line  and a control, do not change when the order of the
columns or rows is changed. 
They are spatially invariant to these permutations.

\subsection{Space filling}
\label{sec:space}

As an aside, it is  of interest to examine the extent to which the
control plots are  scattered across the field on randomization and, thereby,
accommodate regional  heteroscedasticity. This notion can be  quantified by
adopting a suitable space-filling criterion based on the location of the
controls in a given design, and the aggregate distance-based criterion $\phi_2$,
which is a regularized form of the maximin criterion, was chosen for this
purpose \citep{pm:12}. In the present context, the criterion  is defined to be
\[\phi_2(\xi) =\left[
  \sum_{c_a, c_b \in S,\ a<b} \frac{1}{(d_{ab})^2}
  \right ]^{1/2}.\]
Here $\xi$~is a given design with a set $S$  of control cells
labelled~$c_{a}$, $c_b$, \ldots, and
$d_{ab}=||c_a-c_b||$ is the Euclidean distance between cells $c_a$ and $c_b$. 
In accord with
intuition, smaller values of the criterion $\phi_2$ are associated with better
space-filling properties of the controls in the design.  The criterion $\phi_2$
will be used in selected examples later in the text in order to investigate
whether designs obtained by randomization  depend on the initial square
array design. 

\section{Cyclic square array designs}
  \label{sec:cyc}
  \subsection{Nature and properties}

The properties of cyclic square array designs mirror those of the cyclic auxiliary block designs from which they
 are constructed.  The results of this section are therefore drawn  from the
 theory of cyclic designs presented in, for example, the papers by  \citet{dw:65} and \citet{john:66, john:81} and  the book by  \citet{johnw:95}.  More specifically, it follows from the results derived in Section~\ref{sec:metrics} that the structure of the family of auxiliary block designs $\Gamma_c(k,t)$ in terms of equivalence classes and average variance matrics translates immediately to the structure of the corresponding family of square array designs $\Sigma_c(t,k)$.  It is worthwhile exploring this translation a little further.

The development here  hinges on  the fact that the spacing  between the $k$~controls within the first row of a $t \times t$ square array design with $k$ controls $\Sigma_c(t,k)$  is the  same as that of the treatments within
  blocks of the associated cyclic auxiliary block design~$\Gamma_c(k,t)$. Consider first a cyclic square array design $\Sigma_c(t,k)$  specified by the column co\-ordinates
of the first row, say $[j_1, j_2, \ldots, j_k]$, where $j_1 < j_2 <\cdots < j_k$.
Then the spacing between the
$k$~controls in that row, and hence between the controls in each row of the
array, 
is given by the sequence $s_1=j_2-j_1$,
$s_2=j_3-j_2$, \ldots, $s_{k-1}=j_k-j_{k-1}$,
$s_k=(t-j_k)+j_1=t-\sum_{\ell=1}^{k-1} s_{\ell}$, that is $(s_1, \ldots, s_k)$,
which is an integer partition of~$t$. Shifting the column indices of a square
array design cyclically, modulo~$t$, or permuting the rows and columns  does 
 not affect the spacing of the controls. The design can therefore be represented uniquely by a sequence
of spacings $(s_1, \ldots, s_k)$, where $\sum_{\ell=1}^{k} s_{\ell} = t$,
 and is denoted here by $C(s_1, \ldots, s_k)$. 
For ease of interpretation, 
a rule for ordering spacing sequences is adopted so that a single sequence defines a cyclic
set of spacings uniquely.   Specifically, if the sequence
$(s_1, \ldots, s_k)$ occurs first and $(s'_1,\ldots, s_k')$ is another cyclic
ordering of that sequence and these two sequences differ for the first time in
position~$j$, then $s_j < s_j'$ must hold.

The notion of equivalence classes in the context of $t \times t$ cyclic
square array designs with $k$~controls now follows directly from the paper of  \citet{dw:65}.  Consider 
renumbering the rows and columns of the square array as $0$, $1$, \ldots,
$t-1$.  Suppose that a  permutation~$\sigma$ of the set $\{0,1, \ldots, t-1\}$ of the attendant equireplicate cyclic design 
transforms the design specified by $C(s_1, \ldots, s_k)$ into a different design
$C(s_1', \ldots, s_k')$.  Then applying $\sigma$ to the rows and columns of
the square array transforms the square array design defined by
$C(s_1, \ldots, s_k)$ into the one defined  by $C(s'_1, \ldots, s'_k)$.
These two square array designs are  therefore isomorphic and have the same 
values of $A_{ct}$ and $A_{tt}$. The permutation $\sigma$  can be taken from \citet{dw:65} as a permutation of the set $\{0, 1, \ldots, t-1\}$ defined by multiplication by a co-prime~$i$ of~$t$, denoted~$R(t,i)$, or, occasionally, from another form of permutation.

Finally, connectedness of cyclic block designs is discussed by \cite{johnw:95}.
The following theorem generalizes this notion to cyclic square array designs.

\begin{thm}
\label{thm:hcf}
  Consider a $t \times t$ square array design in which each of $k$~controls
  occupies a single left-to-right down-diagonal and the remaining $t(t-k)$ plots
  are occupied by test lines with single replication. Denote the spacings by
  $s_1$, \ldots, $s_k$, as above.  Then there is an unbiased linear estimator
  of the difference between any pair of test lines if and only if
the highest common factor of $s_1$, \ldots, $s_k$ is equal to~$1$.
  \end{thm}

\begin{pf}
  Denote by $q$ the highest common factor of the spacings.
   Since each control $A$, $B$, \ldots, occurs once in each row and once in
  each column, the difference between their average responses gives a linear
  unbiased estimator of the difference between $\tau_A$ and $\tau_B$.  Suppose
  that $A$ and $B$ occur in columns $j$ and $\ell$ of the same row, where
  $j<\ell$.  Then the difference
  between the responses on those two plots gives a linear unbiased estimator
  of $\tau_A -\tau_B +\gamma_j-\gamma_\ell$, and hence of $\gamma_j-\gamma_\ell$.
  Hence  $\gamma_j - \gamma_\ell$ can be estimated whenever $\ell -j$ is a linear
  combination of the spacings $s_1$, \ldots, $s_k$.
  If $q=1$ then  $\ell -j$  can always be expressed as such
  a linear combination.  In this case, a similar argument shows that  all differences 
  of the form $\rho_u-\rho_v$  can be estimated for different rows $u$ and~$v$.

  Suppose that $q=1$.
Let $a$ and $b$ be two different test lines. If they are both in the
  same row, in columns $j$ and~$\ell$, then the difference between their
  responses gives a linear unbiased estimator of $\tau_a -\tau_b + \gamma_j
  -\gamma_\ell$ and hence of $\tau_a-\tau_b$.  A similar argument gives a linear
  unbiased estimator of $\tau_a-\tau_b$ if they are both in the same column.
  If $a$ is in cell $(u,j)$ and $b$ is in cell $(v,\ell)$ then
  $y(u,j)  - y(v,\ell)$ gives a linear unbiased
  estimator for $\tau_a + \rho_u + \gamma_j -\tau_b - \rho_v -\gamma_\ell$
  and hence of $\tau_a - \tau_b$.

  On the other hand, suppose that $q>1$.
  Draw a graph whose vertices are the cells with control
  treatments, joining two vertices if they are in the same row or column.
  This graph is disconnected, having $q$~separate components, so there is no
  way of knowing whether a difference between two components is caused by a
  difference between row effects or a difference between column effects. Hence
  differences between test lines cannot all be estimated.  \qed
\end{pf}

The construction of $t \times t$ cyclic square array designs with
$k$~controls is now straightforward.  Thus, it is possible to completely enumerate all the appropriate  cyclic
auxiliary block designs $\Gamma_c(k,t)$ and, thereby, all the requisite
square array designs $\Sigma_c(t,k)$ for small values of
$t$ and $k$ with $3\leq k<t$.
Equivalence classes can then be identified and
the average variance metrics $A_{ct}$ and $A_{tt}$ calculated from the cyclic design metric $A_c$  by invoking the formulae in Equations~(\ref{eq:att}) and~(\ref{eq:act}). 
As $t$ and $k$ increase,
so complete enumeration becomes prohibitive in terms of
computer time.  However, from a practical perspective,  it is only necessary to obtain a square array design for which the average variance metrics are a minimum. To this end, the initial blocks of equireplicate cyclic designs
which comprise $t$ treatments replicated $r$ times in $b$~blocks of size~$k$
and which  maximize the overall efficiency factor, and hence minimize the average variance metric $A_c$, are readily available from 
papers such as those of \citet{jwh:72} and  \citet{ lh:82} and  from the  package 
CycDesigN \citep{cycdes}. A cyclic design so obtained can immediately be transformed to a $t \times t$ square array design with $k$ controls and minimum values of the average variance metrics $A_{ct}$ and $A_{tt}$ calculated. 

  An example is introduced here in order to  fix ideas. 
 \begin{eg}
\label{eg:kempton}
The percentage of plots occupied by the controls for the $7\times 7$
square array with 3 controls introduced by \cite{fedr:75} is
$42.86$\%, which is, in practical terms, extremely high. 
A more realistic setting which satisfies the proposal of  \citet{kem:84} was 
therefore chosen here, that
of the $12 \times 12$ square array design with $3$~controls, 
where the percentage
of plots occupied by the controls is~$ 25$\%. The $3 \times 12$ cyclic auxiliary block
design and the $12 \times 12$ square array design  with $3$~controls
which it induces are  shown   in Figure 2.
  
The requisite cyclic square array designs of can be identified as partitions of the number of treatments, that is $12$. In addition, the integers $5$, $7$ and $11$ are co-prime to~$12$. Let $\Delta$ be
   the square array design from a design represented by $C(3,4,5)$ shown in
  Figure 2(b) with rows and columns renumbered  as
  $0$, $1$,\ldots, $11$ and the attendant auxiliary block design determined by
  the initial block $\{0,3,7\}$.  Applying the permutation $R(12,5)$ to
  the   rows and columns of $\Delta$ simultaneously
  yields a square array design from the design $C(1,3,8)$. Similarly,
  applying the permutation $R(12,11)$ to the rows and columns of $\Delta$
  simultaneously  reverses the order of the  spacing and yields a square array
  design $C(3,5,4)$. The latter
  operation has a geo\-metric interpretation in that, if a square array design
represented by $C(3,4,5)$ is rotated through $180^{\circ}$, the new
  design is $C(3,5,4)$ and so the spatial configuration, and
  hence the average variance metrics $A_{ct}$ and $A_{tt}$ remain the same.
  Similarly, applying $R(12,11)$ to the design defined by 
  $C(1,1,10)$ reverses the spacings but does not change the design.
  The connected designs can all be so identified by multiplicative permutation. In addition, it follows from Theorem~\ref{thm:hcf}  that the square array designs
 $C(2,2,8)$, $C(3,3,6)$, $C(4,4,4)$, $C(2,4,6)$ and
  $C(2,6,4)$ are not connected. The equivalence classes and average variance metrics of the $12 \times 12$ square array designs with $3$~controls are summarized compactly in
  Table 1. The  results confirm the linear relationships between the metrics $A_{tt}$, $A_{ct}$ and
  $A_c$.
    \end{eg} 
    
 \begin{figure}[ht]
  \caption{ Representation of (a)~a $3\times 12$ cyclic auxiliary block design
  and (b)~the resultant  $12 \times 12$ cyclic square array
    design with $3$~controls.\label{fig:kempt12}}
\begin{center}
  $ \begin{array}{r|c|c|c|c|c|c|c|c|c|c|c|c|}
      \multicolumn{1}{c}{}  & \multicolumn{1}{c}{1} & \multicolumn{1}{c}{2} &
      \multicolumn{1}{c}{3} & \multicolumn{1}{c}{4} & \multicolumn{1}{c}{5}
      & \multicolumn{1}{c}{6} & \multicolumn{1}{c}{7} & \multicolumn{1}{c}{8}
      & \multicolumn{1}{c}{9} & \multicolumn{1}{c}{10} &
      \multicolumn{1}{c}{11} & \multicolumn{1}{c}{12}\\
      \cline{2-13}
      A & 1 & 2 & 3 & 4 & 5 & 6 & 7 & 8 & 9 & 10 &11 & 12\\
      \cline{2-13}
\mbox{(a)}\qquad   B &   4 & 5 & 6 & 7 &8 & 9 & 10 & 11 & 12 & 1 & 2 & 3\\
      \cline{2-13}
      C & 8 & 9 & 10 & 11 & 12 & 1 & 2 & 3 & 4 & 5 &  6 & 7\\
            \cline{2-13}
\multicolumn{1}{c}{}\\
\multicolumn{1}{c}{}   & \multicolumn{1}{c}{1} & \multicolumn{1}{c}{2} &
      \multicolumn{1}{c}{3} & \multicolumn{1}{c}{4} & \multicolumn{1}{c}{5}
      & \multicolumn{1}{c}{6} & \multicolumn{1}{c}{7} & \multicolumn{1}{c}{8}
      & \multicolumn{1}{c}{9} & \multicolumn{1}{c}{10} &
      \multicolumn{1}{c}{11} & \multicolumn{1}{c}{12}\\
      \cline{2-13}
      1  & A & & & B & & & &C & & & &\\
      \cline{2-13}
      2 & &A & &  & B &  & & &C & & &\\
      \cline{2-13}
      3 & & & A & & & B &  &  &  &C & &\\
      \cline{2-13}
      4 & & & &A  & & &B & & &  & C & \\
      \cline{2-13}
      5 &  & & & & A & & & B & && &  C\\
      \cline{2-13}
      \mbox{(b)} \qquad 6 &C &  & & && A & & &B & && \\
      \cline{2-13}
      7 & &C &  & & & & A & & & B & &\\
      \cline{2-13}
      8 & & &C &  & & & & A & & & B & \\
      \cline{2-13}
      9  & & & & C & &  & & & A & & & B\\
      \cline{2-13}
      10 &  B & & & &C & & &  & & A & &\\
\cline{2-13}
11 & & B &  &  & &C & & &  & & A &\\
\cline{2-13}
12 & & &B & & &  &C &  & & & & A\\
\cline{2-13}
      \end{array}$    
\end{center}
  \end{figure}

 \begin{table}[ht]
\caption{ 
The designs representing the equivalence classes of the $3 \times 12$
  cyclic auxiliary block designs and the attendant $12 \times 12$ square array
  designs with $3$~controls, together with 
  the average variance metrics  $A_{c}$, $A_{cc}, A_{ct}$ and $A_{tt}$.
}
\label{tab:des12}
    \centering
{  \addtolength{\tabcolsep}{-1mm}
\begin{tabular}{|l|c|ccc|}
  \hline 
  Designs representative  of & \multicolumn{4}{|c|}{Average variance metrics}\\
the isomorphism classes &  $A_{c}$ & $A_{cc}$ & $A_{ct}$ & $A_{tt}$ \\ 
  \hline
  $C(3,4,5)$, $C(3,5,4)$; $C(1,3,8)$, $C(1,8,3)$ & 
$0.9911$ & $0.1667$  & $2.0910$ & $4.0341$  \\
$C(1,4,7)$, $C(1,7,4)$ & $0.9920$ & 
$0.1667$ &  $2.0921$  & $4.0363$    \\ 
$C(1,2,9)$, $C(1,9,2)$;  $C(2,3,7)$,  $C(2,7,3)$  & $1.0186$ & 
$0.1667$ & $2.1246$ & $4.1020$ \\ 
$C(1,5,6)$, $C(1,6,5)$ & $1.2045$ &
$0.1667$ &  $2.3518$  & $4.5607$ \\
$ C(1,1,10)$, $C(2,5,5)$  & $1.3831$ & 
$0.1667$ & $2.5701$ & $5.0013$ \\
\hline
\end{tabular}}
\end{table}

Finally, it is interesting to assess whether or not randomization of a
square array design yields designs whose space-filling
properties  depend
on the initial design itself. This dependency can best be examined by
simulation and, to illustrate, a study was conducted on the running example
of the $12 \times 12$ square array design with $3$~controls. 
An augmented row-column square array design was selected from each of the
three designs $C(3,4,5)$, $C(1,3,8)$ and $C(1,4,7)$, and permutations of the
ordering $1$, $2$, \ldots, $12$ from the doubly transitive subgroup
$\mathrm{PSL}(2,11)$ of the symmetric group $S_{12}$ were obtained. 
A set of generators for the subgroup was elicited from \citet{gap} and the
elements found using functions in the programming language \citet{Mathematica}.
The permutations were then applied, two at a time, to each of the square array
designs.  The discrete distributions of the $\phi_2$ values for the designs so
generated were all roughly bell-shaped, 
with summary statistics presented in Table 2. It is clear from
these that, from a practical perspective, the space-filling properties of
the  designs generated by permutation do not depend sensitively on the initial
square array design selected.
\begin{table}[ht]
\caption{
Descriptive statistics for the distribution of  $\phi_2$ values
  generated from individual square array designs taken from three designs
 with $t=12$ and $k=3$,
  using permutations from
  the subgroup $\mathrm{PSL}(2,11)$.} 
\label{tab:phip12}
  \centering
\begin{tabular}{|c|cccccc|}
  \hline
 & \multicolumn{6}{|c|}{descriptive statistics for the $\phi_2$ values} \\ 
  Design  &  min & {\footnotesize Q3} & median & mean  & {\footnotesize Q1}
  & max \\
  \hline
$C(3,4,5)$  & 6.042 & 6.717 & 6.845 & 6.855 & 6.983 & 8.069 \\  
$C(1,3,8)$  & 6.113 & 6.729 & 6.850 & 6.855 & 6.976 & 7.984 \\ 
 $C(1,4,7)$ &  6.223 & 6.734 & 6.849 & 6.855 & 6.970 & 8.027  \\
 \hline
\end{tabular}
\end{table}

\subsection{Examples}
\subsubsection{Youden squares as auxiliary block designs}

If the $k \times t$ auxiliary block design is a  Youden square,  it is a symmetric
BIBD and thereby  minimizes the average variance metric $A_{c}$  over all $k \times t$
equireplicate designs. The
resulting $t \times t$ square array design with $k$~controls  therefore minimizes the
metrics  $A_{ct}$ and $A_{tt}$.   Since $A_{abd} = 2k/\lambda t$, where $\lambda=k(k-1)/(t-1)$ is  the number of times any pair of treatments concur in the same block of the Youden square, it follows from Equations~(\ref{eq:att}) and~(\ref{eq:act}) that
\begin{equation}
 A_{ct} = 1 + \frac{1}{t} + \frac{2k}{\lambda t} \mbox{~~and~~}  A_{tt} = 2  + \frac{4k (t-k)}{(t_1-1)\lambda}.
  \label{eq:tt}
  \end{equation}

For practical values of $k$ and $t$, a BIBD
can be found in the tables of \citet[p.~406]{hall:98},
whether or not it is cyclic.
Only a small number of Youden squares exist, however, and, of these, very few
are suitable for use
in the present context.
The numbers of plots and controls available 
are often limited, and,
furthermore, the percentage of plots allocated to controls should fall within,
or close to,  the $20$ to $25$\% window. 
Four
$t \times t$ square array designs with $k$~controls  which can be construed as
being workable within the present framework and which are also generated  from Youden
squares are presented, together with some key properties, in
Table 3, using the results in Theorem~\ref{thm:acc} and
Equations~(\ref{eq:tt}). For completeness, the $7 \times 7$ square array design with 3 controls introduced by \citet{fedr:75} is also included in the table. 
The symmetric BIBD with $t=16$ and $k=6$ is not cyclic
and 
so not relevant 
here, but it does give the Youden
square shown in Figure 3 and
will be used in Section~\ref{sec:other}.
\begin{figure}
\caption{A non-cyclic $6 \times 16$ BIBD,   shown as a Youden square\label{fig:16}}
\begin{center}
  $ \begin{array}{r|c|c|c|c|c|c|c|c|c|c|c|c|c|c|c|c|}
      \multicolumn{1}{c}{}  & \multicolumn{1}{c}{1} & \multicolumn{1}{c}{2} &
      \multicolumn{1}{c}{3} & \multicolumn{1}{c}{4} & \multicolumn{1}{c}{5}
      & \multicolumn{1}{c}{6} & \multicolumn{1}{c}{7} & \multicolumn{1}{c}{8}
      & \multicolumn{1}{c}{9} & \multicolumn{1}{c}{10} & \multicolumn{1}{c}{11}
      & \multicolumn{1}{c}{12}& \multicolumn{1}{c}{13} & \multicolumn{1}{c}{14}
      & \multicolumn{1}{c}{15} & \multicolumn{1}{c}{16}\\
      \cline{2-17}
      A & 2 & 4 & 1 & 3 & 6 & 8 & 5 & 7 & 10 & 12 & 9 & 11 & 14 & 16 & 13 & 15\\
      \cline{2-17}
      B & 3 & 1 & 4 & 2 & 7 & 5 & 8 & 6 & 11 & 9 & 12 & 10 & 15 & 13 & 16 & 14\\
      \cline{2-17}
      C & 4 & 3 & 2 & 1 & 8 & 7 & 6 & 5 & 12 & 11 & 10 & 9 & 16 & 15 & 14 & 13\\
      \cline{2-17}
      D & 5 & 6 & 7 & 8 & 13 & 14 & 15 & 16 & 1 & 2 & 3 & 4 & 9 & 10 & 11 & 12\\
      \cline{2-17}
      E & 9 & 10 & 11 & 12 & 1 & 2 & 3 & 4 & 13 & 14 & 15 & 16 & 5 & 6 & 7 & 8\\
      \cline{2-17}
      F & 13 & 14 & 15 & 16 & 9 & 10 & 11 & 12 & 5 & 6 & 7 & 8 & 1 & 2 & 3 & 4\\
      \cline{2-17}
  \end{array}$
      \end{center}
  \end{figure}

\begin{table}[ht]
\caption{The $t \times t$ square array designs with $k$~controls generated from
  auxiliary block designs which are Youden squares with parameter $\lambda$,
  together with the number of plots required~$t^2$, the percentage of plots
  allocated to controls, an initial block taken from Hall (1998) and the
  average variance metrics $A_c, A_{cc}, A_{ct}$ and $A_{tt}$.}
\label{tab:youden}
  \centering
    {  \addtolength{\tabcolsep}{-1mm}
      \begin{tabular}{|c|c|c|c|c|c|c|ccc|}
  \hline 
  &  &  & number  &  percent & \multicolumn{1}{c|}{initial} &
  \multicolumn{4}{|c|}{Average variance metrics} \\   
  $t$ &  $k$ & $\lambda$ & of plots  & controls & \multicolumn{1}{c|}{block}
    & $A_c$ & $A_{cc}$ & $A_{ct}$ & $A_{tt}$\\\hline
   $ 7 $ & $3$ & $1$ & $~49$ & $42.86$ & $\{1,2,4\}$  & $0.8571$ &   $0.2857$
 & $2.0000$ & $3.7778$   \\ 
  $13$ & $4$ & $1$ & $169$  & $30.77$ & $\{1,2,4,10\}$ & $0.6154$ & $0.1538$ & $1.6923$
  & $3.2414$\\ 
  $16$ & $6$ & $2$ & $256$ & $37.50$ & \mbox{not cyclic} & $0.3750$ & $0.1250$ &
  $1.4375$ & 2.7547\\
  $21$ & $5$ & $1$ & $441$  & $23.81$ & $\{3,6,7,12,14\}$  & $0.4762$ & $0.0952$ & $1.5238$
  & $2.9552$\\ 
  $31$ & $6$ & $1$ & $961$  & $19.35$ & $\{1,5,11,24,25,27\}$ & $0.3871$ & $0.0645$ &
  $1.4194$ & $2.7752$\\ \hline
\end{tabular}}
\end{table}

\subsubsection{Non-Youden equireplicate cyclic designs as auxiliary block
  designs}

The tables of \citet{lh:82} give a wealth of equireplicate cyclic
designs, but some are not useful as auxiliary block designs in the present
context
because they do not satisify the condition that $k/t$
is between $20$\% and $25$\%.
To investigate this restriction in a little more detail, the
$147$~cyclic square array designs with array size~$t$ ranging from
$10$ to~$30$ and number of controls~$k$ from $3$ to~$9$, 
with the ratio $k/t$ falling within the $20$\% to $25$\% and, in addition,
within the $15$\% to $30$\% windows
were identified.
The minimum average variance metrics $A_{tt}$ for the 
designs which comply with these limits were then obtained using the tables
from \citet{lh:82}. 
\begin{table}[!b]
\caption{ The values of the average variance metric $A_{tt}$ for $t \times t$
  cyclic  square array designs with $k$~controls  for which the percentage of
  control plots in the field plan lies between $15$\% and $30$\%, with
  $t$~ranging from $10$ to~$30$ and $k$~from $3$ to~$9$. The entries
  corresponding to designs with a $20$\%--$25$\% window of control plots are
  highlighted in blue.}
\label{tab:bigtab}
  \centering
  \begin{tabular}{r|ccccccc|}
 \multicolumn{7}{c}{\hspace{21mm} Number of Controls $k$} \\
 $t$ & {\footnotesize $3$} & {\footnotesize $4$} & {\footnotesize $5$} &
   {\footnotesize $6$} & {\footnotesize $7$} & {\footnotesize $8$} &
     {\footnotesize $9$} \\ \hline
{\footnotesize $10$} & $3.9636$ &  & & & & & \\ 
{\footnotesize $11$} & $4.0332$ & &  &  & & &  \\ 
{\footnotesize $12$} & \color{blue}{$4.0341$} & & & &  &  &   \\ 
{\footnotesize $13$} & \color{blue}{$4.0465$} & & & &  &  & \\ 
{\footnotesize $14$} & \color{blue}{$4.0901$} & $3.2566$ & & & & & \\ 
{\footnotesize $15$} & \color{blue}{$4.1279$} & $3.2683$ &  &  & & &  \\ 
  {\footnotesize $16$} & $4.1559$ & \color{blue}{$3.2821$} & & & &  &   \\ 
  {\footnotesize $17$} & $4.1893$ & \color{blue}{$3.2935$} & $2.9546$ & & & &
  \\ 
     {\footnotesize $18$} & $4.2273$ & \color{blue}{$3.2997$} & $2.9558$ & & & &
     \\ 
   {\footnotesize $19$} & $4.2250$ & \color{blue}{$3.3053$} & $2.9561$ &  & & &
        \\ 
 {\footnotesize $20$} & $4.2623$ & \color{blue}{$3.3117$} & \color{blue}{$2.9618$}
           & $2.7677$ &  &  &   \\ 
 {\footnotesize $21$} &   & $3.3227$ & \color{blue}{$2.9552$} & $2.7696$
 & & & \\ 
 {\footnotesize $22$} &  & $3.3265$ & \color{blue}{$2.9655$} & $2.7711$
 & & & \\ 
 {\footnotesize $23$} &  & $3.3348$ & \color{blue}{$2.9639$} & $2.7724$
 & & &  \\ 
 {\footnotesize $24$} &  & $3.3377$ & \color{blue}{$2.9670$} &
 \color{blue}{$2.7733$} & $2.6420$ &  &   \\ 
       {\footnotesize $25$} &  & $3.3455$ & \color{blue}{$2.9706$} &
       \color{blue}{$2.7742$} & $2.6435$ & &\\ 
        {\footnotesize $26$} & & $3.3520$ & $2.9741$ & \color{blue}{$2.7748$}
             & $2.6448$ & & \\ 
     {\footnotesize $27$} &   &  & $2.9766$ & \color{blue}{$2.7752$} & $2.6459$
             & $2.5521$ &  \\ 
     {\footnotesize $28$} & & & $2.9798$ & \color{blue}{$2.7756$} &
     \color{blue}{$2.6469$} & $2.5533$ &   \\ 
           {\footnotesize $29$} & &  & $2.9829$ & \color{blue}{$2.7792$} &
           \color{blue}{$2.6479$} & $2.5535$ & \\ 
                 {\footnotesize $30$} & & & $2.9850$ & \color{blue}{$2.7774$} &
                 \color{blue}{$2.6484$} & $2.5545$ & $2.4850$\\ \hline
\end{tabular}
\end{table}
The results are summarized in
Table 4, and should act as a  valuable guide for the
practitioner planning an unreplicated trial. 
It is clear from the table that,
for a given array size~$t$, the value of~$A_{tt}$ decreases with
increasing numbers of controls and, therewith, a decreasing number of test
lines. However, the trend in the average variance metric~$A_{tt}$ for a fixed
number of controls as $t$ increases is more nuanced.
The value of~$A_{tt}$ increases very slightly as $t$~increases and
less so for larger values of $k$. This 
could,
arguably, be attributed to a `saturation' in the number of controls required to
yield precise estimates of pairwise differences of the test-line effects in the
row-column designs. 

For completeness, six specific examples of the $(t,k)$ pairs  for which no
associated cyclic design is a BIBD were considered.
The
requisite designs were derived by complete enumeration, and details of
their properties are presented in Table 5.
The results underscore the fact that the numbers equivalence classes increase rapidly with the number of
treatments and controls, and can readily become challenging to compute.
However, the focus here is on finding  augmented row-column square array
designs which minimize $A_{ct}$ and $A_{tt}$, so that the results of this table,
while appealing in theory, are not needed in practice. In other words, the
average variance metrics $A_{ct}$ and $A_{tt}$ for an individual square array
design can be readily calculated, and then its efficiency relative to a design
with minimum metric values found. 
\begin{table}[!htb]
\caption{The $t \times t$ square array designs with $k$~controls generated from
  cyclic auxiliary block designs which are not Youden squares, together with
  the percentage of control plots, the number of  isomorphism
  classes and the minimum average variance metrics $A_c$, $A_{cc}, A_{ct}$
  and~$A_{tt}$.}
\label{tab:cyclic}
  \centering
    {  \addtolength{\tabcolsep}{-1.2mm}
\begin{tabular}{|c|c|c|c|c|ccc|}
  \hline
  & &   & Numbers of & \multicolumn{4}{|c|}{Minimum}\\
  &    & Percentage &  \multicolumn{1}{|r|}{isomorphism}   &  \multicolumn{4}{|c|}{average variance metrics}  \\   
  $t$ &  ~$k$~  & of controls &  classes & $A_c$ & $A_{cc}$ & $A_{ct}$ & $A_{tt}$\\ \hline
  $9$ &  $3$  & $33.33$ &  $3$ & $0.9229$ & $0.2222$ & $2.0453$ & $3.9037$\\ 
$10$ &  $3$ &  $30.00$ &  $4$ & $0.9527$ & $0.2000$ & $2.0678$ & $3.9636$\\ 
$16$ &  $4$ &  $25.00$ &  $19$ & $0.6352$ & $0.1250$ & $1.7002$ & $3.2821$\\ 
$16$ & $6$ &  $37.50$ &  $64$ & $0.3766$ & $0.1250$ & $1.4399$ & $2.7595$\\
$25$ & $5$ &  $20.00$ &  $110$ & $0.4836$ & $0.0800$ & $1.5243$ & $2.9706$\\ 
$30$ & $6$ & $20.00$ &   $2{,}310$  & $0.3879$ & $0.0667$ & $1.4215$ & $2.7774$\\ \hline
\end{tabular}}
\end{table}

\section{Non-cyclic auxiliary block designs}
\label{sec:other}

The cyclic square array designs of  Section~\ref{sec:cyc}  minimize the average
variance metrics $A_{ct}$ and $A_{tt}$ only over the space of 
cyclic designs.  It is therefore important  to explore the use of  non-cyclic $k \times t$
equireplicate incomplete-block designs 
as auxiliary block designs in the construction of $t \times t$ square array designs with
$k$~controls. 
Thus, equireplicate block designs, and in particular those
which are $A$-optimal over all designs and for which the value of the average
variance metric~$A_{abd}$ is necessarily less than or equal to that of its cyclic design
counterpart, were sought. The design in Figure 3 gives one example.
This strategy resonates with that introduced by \citet{mukerjee:24} in a
study on block designs for one-way heterogeneity in unreplicated trials

The equireplicate cyclic design with $t = 8$ and $k = 3$
and initial block $\{1,2,5\}$ is partially balanced with two associate classes
and the database at \verb+designtheory.org+ indicates that this design is
globally $A$-optimal.
Also, compare the $A_{ct}$ and $A_{tt}$ metrics for the $16 \times 16$ square array
designs with $6$ controls induced by the $6 \times 16$ BIBD and
the optimal
$6 \times 16$ cyclic design
presented in Tables 3 and 5,
 respectively.
Those for the BIBD-induced design are
smaller than those for the cyclic design, in accordance with Section~\ref{sec:metrics}.

The following examples are drawn from the literature.
\begin{eg}
\label{eg:sbdes}
  \citet{baileys:86} introduced the family of rectangular lattices for which 
  $t=b=n(n-1)$ and $k=n-1$, where $n$~is an integer greater
  than~$3$. Setting $n=4$ yields the equireplicate incomplete-block design in
  Table~11 of their paper, and replacing treatment letters by numbers $1$
  to~$12$ yields the auxiliary block design in Figure 1(a),
  which is
  represented as  the $12 \times 12$ row-column square array design with
  $3$~controls in Figure 1(b).
  \end{eg}

\begin{eg}
  \label{eg:CB}
  \citet{chengb:91} showed that square lattice designs, which have $t=k^2$, are $A$-optimal.
  Example~\ref{eg:CB}.1 with
  $k=3$ and Example~\ref{eg:CB}.2 with
  $k=4$ are square lattice designs which  yield the equi\-replicate
  incomplete-block designs presented as auxiliary block designs in
  Figures 4(a) and 4(b), respectively.
  Example~\ref{eg:CB}.3  is  a square lattice design with 
  $k=5$ with potential use here, but the auxiliary block design
  is too large to be represented compactly in the text.
\end{eg}

\begin{eg}
\label{eg:tri}
  The triangular association scheme $T(5)$ has ten elements, consisting of all
  unordered pairs from the set $\{1, 2, 3, 4, 5\}$. Pairs which have a number
  in common are first associates; pairs with no number in common are second
  associates. A partially balanced design with $t = b = 10$ and $k = 3$ can be
  constructed by taking each treatment to define a block which consists of all
  treatments with no number  in common with it. An equi\-replicate
  incomplete-block design can then be obtained by a relabelling of the
  treatments and blocks, giving the auxiliary block design 
  in Figure 4(c). 
  \end{eg}

  The value $A_{abd}$ for the auxiliary block design, and values of the
  average variance metrics $A_{cc}$, $A_{ct}$ and
  $A_{tt}$ of the square array designs in  Examples~\ref{eg:sbdes},
  \ref{eg:CB} and \ref{eg:tri},   are shown in Table 6.
  Comparison of these with those in
  Table 1 and Table 5
confirms the results in Section~\ref{sec:metrics}.
\begin{figure}[ht]
  \caption{Auxiliary block designs  for (a) Example 5.2.1  with $t=9$
  and $k=3$, (b) Example 5.2.2 with $t=16$ and $k=4$,
  and (c) Example 5.3 with $t=10$ and $k=3$.
\label{fig:aopt1}}
\begin{center}
  $ \begin{array}{r|c|c|c|c|c|c|c|c|c|}
      \multicolumn{1}{c}{}  & \multicolumn{1}{c}{1} & \multicolumn{1}{c}{2} &
      \multicolumn{1}{c}{3} & \multicolumn{1}{c}{4} & \multicolumn{1}{c}{5}
      & \multicolumn{1}{c}{6} & \multicolumn{1}{c}{7} & \multicolumn{1}{c}{8}
      & \multicolumn{1}{c}{9} \\
      \cline{2-10}
      A & 1 & 2 & 3 & 4 & 5 & 6 & 7 & 8 & 9\\
      \cline{2-10}
\mbox{(a)}\quad   B &   4 & 8 & 9 & 3 &1 & 7 & 5 & 6 & 2\\
      \cline{2-10}
      C & 7 & 5 & 6 & 8 & 9 &  2 & 3 & 1 & 4\\
      \cline{2-10}
  \end{array}$
    \end{center}
\begin{center}
  $ \begin{array}{r|c|c|c|c|c|c|c|c|c|c|c|c|c|c|c|c|}
      \multicolumn{1}{c}{}  & \multicolumn{1}{c}{1} & \multicolumn{1}{c}{2} &
      \multicolumn{1}{c}{3} & \multicolumn{1}{c}{4} & \multicolumn{1}{c}{5}
      & \multicolumn{1}{c}{6} & \multicolumn{1}{c}{7} & \multicolumn{1}{c}{8}
      & \multicolumn{1}{c}{9} & \multicolumn{1}{c}{10} &
      \multicolumn{1}{c}{11} & \multicolumn{1}{c}{12} &
      \multicolumn{1}{c}{13} & \multicolumn{1}{c}{14}&
        \multicolumn{1}{c}{15} & \multicolumn{1}{c}{16}\\
        \cline{2-17}
        A & 1 & 8 & 10 & 15 & 9 & 2 & 7 & 16 & 6 & 12 & 3&13 & 14 & 11 & 5 & 4\\
        \cline{2-17}
        \mbox{(b)} \quad
        B & 4 & 5 & 11 & 14 & 1 & 10 & 15 & 8 & 16 & 2 & 9 & 7 & 12 & 13 & 3&6\\
        \cline{2-17}
        C & 3 & 6 &12 &13 & 5 &14 & 11 & 4 & 1 & 15 &  8 & 10 & 7 & 2 & 16 & 9\\
        \cline{2-17}
        D & 2 & 7 & 9 & 16 & 13 & 6 & 3 & 12 & 11 & 5 & 14 & 4 & 1 & 8&10 & 15\\
        \cline{2-17}
  \end{array}$
      \end{center}
\begin{center}
  $ \begin{array}{r|c|c|c|c|c|c|c|c|c|c|}
         \multicolumn{1}{c}{}  & \multicolumn{1}{c}{1} & \multicolumn{1}{c}{2} &
      \multicolumn{1}{c}{3} & \multicolumn{1}{c}{4} & \multicolumn{1}{c}{5}
      & \multicolumn{1}{c}{6} & \multicolumn{1}{c}{7} & \multicolumn{1}{c}{8}
      & \multicolumn{1}{c}{9} & \multicolumn{1}{c}{10}\\
      \cline{2-11}
      A & 1 & 8 & 4 & 5 & 10 & 3 & 2 & 9 & 7 & 6\\
      \cline{2-11}
       \mbox{(c)} \quad
      B & 2 & 9 & 7 & 6 & 3 & 1 & 8 & 4 & 5 & 10\\
      \cline{2-11}
      C & 5 & 10 & 1 & 8 & 4 & 6 & 3 & 2 & 9 & 7\\
      \cline{2-11}
  \end{array}$
\end{center}
  \end{figure}

\begin{table}[ht]
    \caption{Average variance metrics 
      for the
  non-cyclic $k \times t$ auxiliary block designs and the attendant
   square array designs   in
  Examples 5.1, 5.2 and 5.3.}
\label{tab:aopt}
\centering
\begin{tabular}{|cl|c|c|ccc|}
  \hline 
 & &  &  \multicolumn{4}{|c|}{Average variance metrics}   \\
$t$ &  $k$  & Example & $A_{abd}$  &  $A_{cc}$ & $A_{ct}$ & $A_{tt}$ \\ 
  \hline   
9 & 3 & 5.2.1 & $0.916 7$ & $0.2222$ &$2.0370$ & $3.8868$ \\ \hline

10 & 3 & 5.3 & $ 0.9500$ &  $0.2000$ & $2.0643$ & $3.9565$ \\ \hline 
12 & 3 & 5.1 & $0.9803$ & $0.1667$ &  $2.0778$ & $4.0075$ \\ \hline 
16 & 4 & 5.2.2 & $ 0.6333 $ & $0.1250$ &  $1.6979$ & $3.2775$ \\ \hline 
25 & 5 & 5.2.3 & $ 0.4833 $ & $0.0800$ &  $1.5240$ & $2.9699$ \\ \hline 
\end{tabular}
\end{table}

\section{Conclusions}
\label{sec:end}

The main results of in this paper concern the use of 
$k \times t$ auxiliary block designs to construct
augmented row-column $t \times t$ square array 
designs with $k$~controls for unreplicated trials. The construction
is straightforward and allows for considerable
flexibility in terms of numbers of plots,  controls and test lines.
An extensive family of square array designs can be built from
equireplicate cyclic designs, and the properties of these designs, including
spacing, connectedness, categorization by equivalence classes,
and randomization are explored. 
Non-cyclic equireplicate incomplete-block
designs
are also introduced and provide valuable auxiliary
block designs for square array designs. 
Workable 
$t \times t$ 
square array designs with $k$~controls,
that is,  designs which minimize the average   variances of estimates of the
pairwise differences between test-line effects and which have a percentage of
control plots of between $15$\% and $30$\%,  are identified and recommended.  

There is scope for further research. 
The augmented row-column designs  presented here are appropriate only for square
fields, so it would  be interesting to extend the investigation to
rectangular fields.  For example, an $r \times t$  partial latin rectangle with
$r < t$ and $k$~controls, such as the $6 \times 8$ design with $3$~controls
shown by \citet[p.~41]{fedc:05},
can be constructed by deleting rows from a $t \times t$ square array design
with $k$~controls and may somtimes inherit properties, such as connectedness, from
the parent design. 

The notion of the square array designs 
was first
introduced by \citet{fedr:75} but the structure was not pursued until now. It
may therefore be worthwhile to revisit the augmented row-column designs
introduced in the early papers of  \citet{fedr:75} and \cite{fnr:75} from both
a theoretical and a practical perspective. For example, \citet[p.~368]{fnr:75}
introduced a design in which the left-to-right diagonals of a square array are
taken to be transversals, that is designs in which the controls occur exactly
once on each diagonal, but the family of such designs has not been examined
further.

\paragraph{Supplementary Material} The R package {\tt gunrep} and the accompanying programs can be found on GitHub under https://github.com/LindaHaines/gunrep and are subject to {\em caveat utilitor}: user beware.

\end{document}